%% file: main.tex
\documentclass[manuscript,screen,acmsmall ]{acmart}
\AtBeginDocument{%
  \providecommand\BibTeX{{%
    \normalfont B\kern-0.5em{\scshape i\kern-0.25em b}\kern-0.8em\TeX}}}

\usepackage{booktabs}
\usepackage{graphicx}
\setcopyright{acmlicensed}
\copyrightyear{2024}
\acmYear{2024}
\acmDOI{XXXXXXX.XXXXXXX}

\acmConference[Conference acronym 'XX]{Make sure to enter the correct
  conference title from your rights confirmation emai}{June 03--05,
  2018}{Woodstock, NY}
\acmBooktitle{Woodstock '18: ACM Symposium on Neural Gaze Detection,
 June 03--05, 2018, Woodstock, NY} 
\acmISBN{978-1-4503-XXXX-X/18/06}

\definecolor{cBLUE}{HTML}{3282B8}
\definecolor{cGREEN}{HTML}{60A561}
\definecolor{cORANGE}{HTML}{FA824C}
\definecolor{cYELLOW}{HTML}{f0C808}
\definecolor{cLightGrey}{HTML}{CECECE}
\definecolor{cRED}{HTML}{ED1B23}

\newcommand{\weidan}[1]{{\bf \color{purple} [weidan: #1]}}

\newcommand{\reyna}[1]{{\bf \color{cGREEN} [reyna: #1]}}

\newcommand{\ziqi}[1]{\textcolor{violet}{[ziqi: #1]}}

\newcommand\pquote[2]{{``\textit{#2}'' (\textbf{#1})}}

\begin{document}

\title[Cancer Treatment-Induced Cardiotoxicity]{Clinical Challenges and AI Opportunities in Decision-Making for Cancer Treatment-Induced Cardiotoxicity}

\author{Siyi Wu}
\authornote{Both authors contributed equally to the paper}
\affiliation{
    \institution{University of Toronto}
    \city{Toronto}
    \country{Canada}}
\email{reyna.wu@mail.utoronto.ca}

\author{Weidan Cao}
\authornotemark[1]
\affiliation{
    \institution{Ohio State University}
    \city{Columbus}
    \country{USA}}
\email{weidan.cao@osumc.edu}

\author{Shihan Fu}
\affiliation{
    \institution{Northeastern University}
    \city{Boston}
    \country{USA}}
\email{fu.shiha@northeastern.edu}

\author{Bingsheng Yao}
\affiliation{
    \institution{Northeastern University}
    \city{Boston}
    \country{USA}}
\email{fb.yao@northeastern.edu}

\author{Ziqi Yang}
\affiliation{
    \institution{Northeastern University}
    \city{Boston}
    \country{USA}}
\email{ziq.yang@northeastern.edu}

\author{Changchang Yin}
\affiliation{
    \institution{Ohio State University}
    \city{Columbus}
    \country{USA}}
\email{yin.731@buckeyemail.osu.edu}

\author{Varun Mishra}
\affiliation{
    \institution{Northeastern University}
    \city{Boston}
    \country{USA}}
\email{v.mishra@northeastern.edu}

\author{Daniel Addison}
\affiliation{
    \institution{Ohio State University}
    \city{Columbus}
    \country{USA}}
\email{daniel.addison@osumc.edu}

\author{Ping Zhang}
\affiliation{
    \institution{Ohio State University}
    \city{Columbus}
    \country{USA}}
\email{zhang.10631@osu.edu}

\author{Dakuo Wang}
\affiliation{
    \institution{Northeastern University}
    \city{Boston}
    \country{USA}}
\email{d.wang@northeastern.edu}

\
\renewcommand{\shortauthors}{Wu and Cao, et al.}
\newcommand{\projectname}{CardioGuard}
\begin{abstract}

Cardiotoxicity induced by cancer treatment has become a major clinical concern, affecting the long-term survival and quality of life of cancer patients. 
Effective clinical decision-making, including the detection of cancer treatment-induced cardiotoxicity and the monitoring of associated symptoms, remains a challenging task for clinicians.
This study investigates the current practices and needs of clinicians in the clinical decision making of cancer treatment-induced cardiotoxicity and explores the potential of digital health technologies to support this process. 
Through semi-structured interviews with seven clinical experts, we identify a three-step decision-making paradigm: 1) symptom identification, 2) diagnostic testing and specialist collaboration, and 3) clinical decision-making and intervention. 
Our findings highlight the difficulties of diagnosing cardiotoxicity (absence of  unified protocols and high variability in symptoms) and monitoring patient symptoms (lacking accurate and timely patient self-reported symptoms). The clinicians also expressed their need for effective early detection tools that can integrate remote patient monitoring capabilities. 
Based on these insights, we discuss 
the importance of understanding the dynamic nature of clinical workflows, and the design considerations for future digital tools to support cancer-treatment-induced cardiotoxicity decision-making. 

\end{abstract}




\received{20 February 2007}
\received[revised]{12 March 2009}
\received[accepted]{5 June 2009}

\maketitle

\input{sections/1-introduction}

\input{sections/2-related-work}
\input{sections/3-methods}

\input{sections/4-results}

\input{sections/5-discussion}

\input{sections/6-conclusion}


\bibliographystyle{ACM-Reference-Format}
\bibliography{references, reference-ai-technology, reference-digital-technology}


\end{document}

%% file: sections/1-introduction.tex
\section{Introduction}

Cancer has become the leading cause of death worldwide, posing significant challenges to global public health efforts~\cite{bray2021ever}. 
Traditional cancer treatments, such as chemotherapy, have been widely used but often lead to severe toxicity in vital organs (i.e., the liver and kidneys), potentially resulting in life-threatening complications~\cite{malyszko2020link, lefebvre2017kidney, munoz2017radiation, shanholtz2001acute}. 
In response to these challenges, there has been a significant shift towards the development and adoption of new targeted, biologic, and immune-based cancer therapies over the past two decades, with more than $180$ new drug approvals since 2010 alone~\cite{us_fda_drugs}. 
These novel therapies aim to improve therapeutic outcomes by specifically targeting cancer cells and minimizing toxicity to healthy tissues, thereby improving patient quality of life~\cite{hu2024sequential, van2023blinatumomab, wang2013targeting, raymond2011sunitinib, schmid2018atezolizumab, borghaei2015nivolumab, hodi2010improved}. 
In addition, the continuous development of existing therapies, such as chemotherapeutic agents, hormone therapies, and radiation treatments, has also contributed to the increase in survival rates and the reduction of immediate toxic effects~\cite{truong2014chemotherapy, shaikh2012chemotherapy, florescu2013chemotherapy}. 
However, these advancements, along with prolonged patient life, have also revealed new challenges, such as cancer treatment-induced cardiotoxicity, which had previously been overlooked ~\cite{truong2014chemotherapy, shaikh2012chemotherapy, florescu2013chemotherapy}.

\textbf{Cancer treatment-induced cardiotoxicity}, which encompasses a wide spectrum of concurrent cardiovascular diseases (CVDs) caused by cancer treatments, has attracted the attention of clinical research in recent years~\cite{alvi2019cardiovascular, baptiste2019high}.
For certain cancers and treatments, cardiotoxicity could account for up to 30\% of the cardiovascular complications observed in cancer patients~\cite{siegel2022cancer,addison2023cardiovascular,baptiste2019high}. These complications can seriously affect both long-term survival rates and quality of life of cancer patients~\cite{fung2015cardiovascular, youn2014long, daher2012prevention, kourek2022cardioprotective}. Among breast cancer patients, for instance, cancer treatment-induced cardiotoxicity has emerged as the leading cause of morbidity and mortality during long-term follow-up~\cite{henry2018cardiotoxicity}.


Despite the increasing amount of clinical evidence and observation on treatment-induced cardiotoxicity in cancer patients, there is a lack of unified practice to facilitate the systematically clinical decision-making of such clinical severity, including the detection of cardiotoxicity and the monitoring of associated symptoms~\cite{moslehi2018increased,jurcut2008detection}. 
Federal initiatives have been taken to urgently call for collaborative research to identify patients at risk of developing cancer treatment-induced cardiotoxicity, such as proposing known or emerging methods to manage CVDs and optimize outcomes among cancer patients~\cite{NIH_CA_22_001}.
Despite these efforts, effective monitoring, diagnosis, and interventions for treatment-induced cardiotoxicity and the resulting CVDs remain a highly challenging task in clinical practice, particularly after long-term cancer treatment when patients are discharged from hospital care~\cite{barac2015cardiovascular, jurcut2008detection, zheng2018breakthroughs}.
The onset of CVDs due to cardiotoxicity can occur anywhere from a few weeks to more than ten years after cancer treatment~\cite{meacham2010cardiovascular}, complicating long-term patient care.
Patients often lack the necessary health literacy and awareness to recognize early cardiotoxicity symptoms, and clinicians face difficulties implementing effective monitoring strategies for clinical decision making beyond the hospital setting~\cite{clark2019cardiotoxicity}. 

In response to these challenges, various efforts, primarily with patient-centered approaches, have been made to engage cancer patients in proactive monitoring their health and managing potential risks~\cite{larsen2017cardio, cheng2017autoimmune, padegimas2020cardioprotective}. 
However, a critical obstacle remains: there is a lack of understanding of how clinical experts identify cancer treatment-induced cardiotoxicity and monitor associated symptoms within their current workflows.
The lack of understanding of the specific needs and challenges for effective clinical decision-making hinders the development of comprehensive strategies to prevent CVDs and mitigate life-threatening risks for cancer survivors.

The increasing development of technology-integrated tools in diverse clinical decision-making scenarios sheds light on new opportunities to support clinicians in more effective clinical decision making of cancer treatment-induced cardiotoxicity. 
Digital health technologies, such as mHealth applications, wearable devices, smart voice assistants, and predictive AI models, have demonstrated potential in cancer care and other high-stakes clinical scenarios~\cite{clauser2011improving}. 
For example, wearables and mHealth applications can track cancer patients' activities and symptoms, while AI-powered conversational agents have facilitated public and mental health interventions~\cite{hardcastle2020fitbit, qiu2021NurseAMIE, xu2021chatbot, piau2019smartphone, yang2023talk2care, kim2024mindfuldiary}. 
In addition, machine learning models are increasingly being leveraged for collaborative clinical decision making between humans and AI~\cite{sepsis, yagi2024artificial, changArtificialIntelligenceApproach2022}. 
To secure the effectiveness and usability of novel technology-supported tools in clinical experts' daily workflow without causing additional complications and burdens, the technology design and development should be grounded in the current workflow of clinical experts, as well as taking into account the perspectives and concerns of clinicians regarding technology adoption. 
Specifically, in the context of clinical decision making related to cancer treatment-induced cardiotoxicity, it is critical to investigate the existing clinical workflow, understand the comprehensive decision-making process, and identify when and how clinicians need assistance. By aligning technological solutions with clinician needs and workflows, we build a responsible and clinician-centered approach to assist clinical decision making for cancer treatment-induced cardiotoxicity.

To this end, we proposed the following research questions (RQs) in our study to address the aforementioned critical challenges:
\begin{itemize}
    \item RQ1: What are the current practices and needs that clinicians face in cancer treatment-induced cardiotoxicity decision making?
    \item RQ2: How do clinicians envision the opportunity and concern of digital health technologies in the clinical process of cancer treatment-induced cardiotoxicity decision making?
\end{itemize}

In this work, we employed a semi-structured interview approach with seven clinical experts who have extensive experience in different types of cancer and treatment. We probed into their current practices and needs in the current workflow of cancer treatment-induced cardiotoxicity decision making, and asked clinicians to envision and share their perspectives on how digital health technologies could support them in their current workflow.

We discovered that the current practice of managing cardiotoxicity risk induced by cancer treatment involves a complex landscape, where we organize the entire decision making process into a three-step paradigm: \textit{(1)} Symptom Identification; \textit{(2)} Diagnostic Tests and Collaboration among Specialists; \textit{(3)} Clinical Decision-Making and Intervention. 
Critically, we identified the clinical process as a dynamic, multi-stakeholder collaborative decision-making process, generally involving various specialists and requiring careful balancing of clinical decisions. The clinicians highlighted the need for early detection and shared their perspectives on remote monitoring and digital health technologies as valuable tools to support their workflows.
Based on these insights, we propose design considerations for future technology-supported tools to support clinical collaboration and decision-making of cancer-treatment-induced cardiotoxicity.
We emphasize the importance of understanding collaborative efforts in high-stakes, high-uncertainty clinical scenarios, where the evolving nature of diseases and treatments adds complexity to decision-making workflows.

Our work makes contributions in the following aspects: 
\begin{itemize}
    \item We present key challenges in the current practices of diagnosing and monitoring cardiotoxicity risks related to cancer treatment from clinicians' perspectives.
    \item We explore the potential and concerns of leveraging smart technologies to facilitate clinical needs and improve cardiotoxicity risk management in cancer treatment.
    \item We provide design considerations for the future development of technology-supported clinical workflows for cancer patients' cardiotoxicity risk management .
\end{itemize}

%% file: sections/2-related-work.tex
\section{Related Work}
\subsection{Cardiotoxicity from Cancer Treatments}
Cancer has emerged as the leading cause of premature death worldwide~\cite{bray2021ever}.
Advances in cancer therapeutics have led to dramatic improvements in survival, now inclusive of over 19 million patients in the US alone~\cite{siegel2022cancer,addison2023cardiovascular}. 
The primary goal of cancer treatment is to eliminate and prevent the recurrence of cancer, thereby extending the patient's life~\cite{siegel2012cancer}. 
This objective has driven significant advancements in cancer therapies, such as chemo-, targeted-, biologic, and immune-based therapies, and contributed to achieving remission and potential cures~\cite{broder2008chemotherapy, dent2015cancer}.
As a result, cancer mortality rates have significantly declined over the past two decades~\cite{jemal2010declining, howlader2010improved, jemal2005trends}.
However, such cancer treatment options have been proven to impact patients' cardiac health, and the gains in cancer survival have been tempered by a parallel rise in life-threatening cardiovascular toxicity associated with the expanding use of novel cancer treatments~\cite{siegel2012cancer}.

Cardiotoxicity, broadly defined as heart damage resulting from cancer treatment ~\cite{shelburne2014cancer}, is a significant factor affecting the life quality and prognosis of cancer patients~\cite{armenian2017prevention, meacham2010cardiovascular, armenian2018cardiovascular} and the leading cause of treatment-associated morbidity and mortality among cancer survivors~\cite{mertens2001late, armstrong2016reduction,abdel2017population,alexandre2020cardiovascular}. In adult patients, the incidence of cardiotoxicity can be as high as 30\% for certain cardiac conditions~\cite{shelburne2014cancer,yeh2009cardiovascular}. The initial manifestation of cardiotoxicity can remain asymptomatic and then eventually present through different clinical manifestations, including left ventricular (LV) dysfunction, heart failure (HF), arrhythmias, and hypertension~\cite{lenneman2016cardio, chen2012incidence, bowles2012risk}. 
Moreover, evidence indicates that the risk of cardiotoxicity can continue to increase for many years even after the completion of cancer treatment~\cite{meacham2010cardiovascular}.

Due to its complexity, cardiotoxicity introduces significant uncertainty and risk in cancer treatment. Concerns about potential cardiotoxicity can limit the use of optimal cancer therapies, potentially compromising treatment outcomes ~\cite{shelburne2019changing, shelburne2014cancer}. Consequently, clinicians face various challenges in balancing effective cancer treatment and cardiac risk assessment in treatment decision-making and managing long-term cardiac risks and effects of treatment ~\cite{curigliano2016cardiotoxicity, herrmann2020adverse, shelburne2014cancer}. Thus, identifying new ways to effectively improve pre-treatment prediction, and the early delivery of post-treatment intervention to mitigate cardiotoxicity could dramatically improve outcomes among cancer patients faced with the uncertain future.

\subsection{Cardiotoxicity Risk Management and Challenges}
Cardiotoxicity risk management in cancer treatment is crucial due to the increasing prevalence of cardiovascular dysfunction associated with anticancer therapies. 
Strategies are developed with a patient-centered approach to manage cardiotoxicity, focusing on prevention, detection, and monitoring ~\cite{jurcut2008detection,mondal2019cardiotoxicity}.

The initial measure to address this issue is prevention, which encompasses patient monitoring and providing educational resources to those at higher risk. Individuals undergoing cancer treatment frequently prioritize the eradication of cancer, sometimes overlooking the significance of managing hypertension, heart failure, and ischemic symptoms~\cite{chen2015cardiovascular}. Available data suggests that many cardiac may be under-recognized or missed until more severe phenotypes of cardiovascular diseases occur~\cite{bonsu2020reporting}. 
The timely detection of cardiac issues plays a vital role in determining the optimal timing of intervention, which is crucial for promoting cardiac recovery~\cite{cardinale2010anthracycline}. 
Clinicians typically identify potential cardiotoxicity symptoms through diagnostic tests such as electrocardiograms (ECGs) and echocardiograms ~\cite{clark2019cardiotoxicity}. 
Once cardiotoxicity is suspected, patients may be referred to specialists, such as cardiologists, for cardiotoxicity treatment and monitoring. 
At last, active monitoring is needed before, during, and following the cancer treatment process~\cite{pavo2015cardiovascular, negishi2013independent, ky2014early}.
Monitoring cardiac function for early detection of cardiotoxicity is essential, especially for cancer survivors exposed to cardiotoxic treatments~\cite{jurcut2008detection, lipshultz2013long, truong2014chemotherapy}.

While these patient-centered approaches to cardiotoxicity management have been proposed and recognized, they still pose challenges to clinicians. The course of evaluation and follow-up for cardiotoxicity has not yet been clearly defined by clinical guidelines~\cite{jurcut2008detection}.
Additionally, the dynamics of cancer treatment and cardiotoxicity add barriers to achieving a balanced approach to patient care.

\subsection{Digital Health Technologies for Cancer and Cardiotoxicity Risk Management}
The rise of digital health technologies marks a pivotal shift in treating and monitoring cancer and its side effects, integrating tools such as wearable devices (WDs), conversational agents (CAs), and AI models to revolutionize patient care~\cite{gresham2018wearable, cox2018use, xu2021chatbot, changArtificialIntelligenceApproach2022}. 
To offer continuous, real-time, and remote monitoring, WDs have been adopted to capture a wide range of patient health data ~\cite{hardcastle2020fitbit, gresham2018wearable, cox2018use}. 
These include fitness trackers and smartwatches like the Fitbit and Garmin watch, which could monitor physical activity levels and heart rate~\cite{hardcastle2020fitbit, tadas2023using}. 
Advanced multisensory devices track a broader range of physiological parameters, such as oxygen saturation and respiratory rates; some systems leverage those data to analyze sleep patterns and circadian rhythms,~\cite{lachenmeier2022home, gresham2018wearable, cox2018use}. 

Meanwhile, communication technologies, such as telehealth or CAs, gather qualitative patient health data such as daily symptoms for cancer care~\cite{larson2018effect}. Telehealth platforms have facilitated cancer patients' access to healthcare consultations and personalized coaching, boosting patient engagement and adherence to treatment protocols ~\cite{larson2018effect, cox2017cancer, chen2018effect}. 
For instance, the Nurse AMIE project utilizes smart speakers to deliver mental and informational care interventions for women with metastatic breast cancer~\cite{qiu2021NurseAMIE}; \citet{gregory2023exploring} designed a mobile health application for tracking cancer patients' cardiac symptoms via questionnaires.
Yet these technologies may not adhere to the varying, personalized, and urgent needs of cardiotoxicity risk management or encounter acceptance issues within patient populations~\cite{hardcastle2018acceptability, nguyen2017patient}.

The recent advances in AI shed light on more accessible, powerful, and efficient digital solutions to manage cancer patients.
In particular, Large Language Models (LLMs), a deep learning algorithm that generates and understands free-form natural language, have facilitated healthcare communication for various populations~\cite{xu2021chatbot, kim2024mindfuldiary, ma2024evaluating, cuadra2024digital}.
Researchers have developed LLM-powered CAs for cancer patients' diagnosis, decision-making, education, and symptom monitoring ~\cite{xu2021chatbot,piau2019smartphone}. Meanwhile, clinicians also appreciate the adoption of AI predictive modeling in supporting their clinical decision-making, which has been popular research and practical focus among AI researchers, HCI researchers, and practitioners~\cite{changArtificialIntelligenceApproach2022, lalArtificialIntelligenceRisk2023, ahmedAdvancementsCardiotoxicityDetection2024, kwanMultimodalityAdvancedCardiovascular2022, yagiArtificialIntelligenceenabledPrediction2024}. For example, \citet{changArtificialIntelligenceApproach2022} trained AI with clinical, chemotherapy, and echocardiographic parameters to predict cancer treatment-induced cardiac dysfunction (CTRCD) and heart failure, reaching higher accuracy than traditional prediction models; \citet{yagiArtificialIntelligenceenabledPrediction2024}'s model robustly stratified CTRCD risk from baseline electrocardiograms (ECG). 
However, these studies either rely on clinical results for risk prediction or only serve general-purpose communication, while facing reliability, accuracy, or privacy concerns towards AI applications in healthcare~\cite{changArtificialIntelligenceApproach2022, lalArtificialIntelligenceRisk2023, ahmedAdvancementsCardiotoxicityDetection2024}.
Despite the significant potential of digital health technologies to remotely monitor cancer survivors, there is limited knowledge on how these advances could benefit care providers in managing long-term cardiotoxicity risk. To address this gap, we begin by understanding the needs, challenges, and perceptions of cancer treatment clinicians regarding these innovative technologies.

%% file: sections/3-methods.tex
\section{Methods}
Clinical observations, according to existing literature, have identified the increasing significance of cardiotoxicity in cancer patients.
Nevertheless, there is a lack of a comprehensive understanding of current clinical practice in cardiotoxicity encounters and risk management, which, in turn, also potentially hinders the development of clinical guidelines and adoption of technology to facilitate clinical decision-making.
To bridge the gap, we first designed and conducted semi-structured interviews with seven clinical experts in this field, following the established Human-Centered AI research design principles~\cite{
amershi2019guidelines, shneiderman2022human, wang2020human, wang2021brilliant, wang2019human, wang2021cass}, to better understand the practices, needs and challenges.
This study comprises bi-fold purposes: 
\begin{itemize}
    \item to gather insights into the clinicians' current practices and challenges in cardiotoxicity risk management for cancer patients, and,
    \item to explore the potential of smart technologies, such as telehealth devices coupled with AI assistance, as well as clinicians' perspectives and expectations toward the adoption of such technologies, in facilitating the clinical practice of cardiotoxicity risk management. 
\end{itemize}


\subsection{Participants}

We recruited seven clinicians whose daily workflow encompasses the responsibility of different cancer-related monitoring, diagnosis, and treatments.
We expect these cancer experts, with different clinical expertise and from different departments, will directly face cancer treatment-induced cardiotoxicity cases in their day-to-day work.  
Using convenience sampling, we recruited participants by reaching out to colleagues and connections within related fields. Table~\ref{tab:interviewees} presents the demographics of the interview participants, including gender, department, job titles, and year of practice.
We randomly assign a numbered index and use $P\#$ to refer to each participant hereinafter. The interviews were conducted remotely via Zoom, with each session lasting approximately 40 minutes. 
This research was approved by the Institutional Review Board (IRB) of the first author's institution.

\begin{table}[htbp]
    \centering
    \resizebox{\textwidth}{!}{
    \begin{tabular}{c|c|c|c|c}
    \toprule
      P\# & Gender & Department & Job Title & Year of Practice\\
    \midrule
      P1  &  Female & Medical Oncology  & Breast Medical Oncologist & 2 years\\
      P2  &  Male & Radiation Oncology & Gastrointestinal Radiation Oncologist & 6 years\\
      P3  &  Female & Bone and Marrow Transplant and Cellular Therapy & Hematologist& 1.5 years\\
      P4  &  Female &  Internal Medicine    &  Oncologist & 13 years\\
      P5  &  Female & Sarcoma Center    & Sarcoma Medical Oncologist & 2 years\\
      P6  &  Female & Thoracic and Geriatric Oncology   & Thoracic and Geriatric Oncologist & 7 years\\
      P7  &  Male & Internal Medicine   & Cardiologist &  8 years\\
    \bottomrule
    \end{tabular}
    }
    \caption{Demographics of Physician Participants}
    \label{tab:interviewees}
\end{table}

\subsection{Procedures} 
During each interview, one researcher led the interview, while at least one other researcher in our team attended the meeting to take notes and ask clarifying questions.
We acquire the content of each participant before the study.
Our interview design focused on two main aspects. 

First, participants were asked to introduce their background, including their clinical expertise, year of practice, their daily clinical workflow, and whether they had encountered cardiotoxicity cases during their work. 
They were then asked to recall a recent cardiotoxicity encounter during or after cancer treatment. 
Through in-depth conversations and follow-up questions about their cancer treatment-induced cardiotoxicity encounters, we probe into detailed clinical practices for cancer treatments, cardiotoxicity risk management, technologies currently used, and, more importantly, their perceived needs and challenges.

Subsequently, we gathered participants' perspectives as clinical experts on opportunities to take advantage of different types of smart technologies in cardiotoxicity risk management to support their clinical workflow.
Follow-up questions emphasize exploring their perspectives and concerns regarding the potential usage of smart technologies, including the expected information that they believe could be beneficial to support their clinical decision-making and outcomes.

\subsection{Data Analysis}
All interview sessions were recorded with the participants' permission and then transcribed. 
Two researchers initially familiarized themselves with the transcripts to establish a comprehensive understanding, then independently coded the data using an open coding approach~\cite{corbin2014basics}. 
Through an inductive approach, themes and sub-themes were constructed and iteratively refined by assigning codes to the interview transcriptions. 
The similar codes were then organized into higher-level themes.
To ensure the validity and reliability of the coding results, researchers cross-validated their coding, discussed potential disagreements, and established consensus on all codes through discussions following the independent coding process.
Any further discrepancies were resolved through additional discussions with the entire research team.

%% file: sections/4-results.tex
\section{Findings}
\label{sec:findings}

In this section, we present our findings based on the qualitative analysis of the semi-structured interviews with clinical experts. 
In Section~\ref{sec:findings-practice}, we describe current clinical practices for diagnosing and monitoring cancer treatment-induced cardiotoxicity, from symptom identification to diagnostic testing to clinical decision-making. 
Then, in Section~\ref{sec:findings-challenges} and \ref{sec:findings-needs}, we illustrate challenges in current practices in cardiotoxicity risk management and explore the participants' perspective toward the potential of digital health technologies in addressing these challenges and facilitating clinical decision-making.

\subsection{Current Clinical Practices of Cancer Treatment-Induced Cardiotoxicity Risk Management}
\label{sec:findings-practice}

\begin{figure*}[!tp]
    \centering
    \includegraphics[width=1\textwidth]{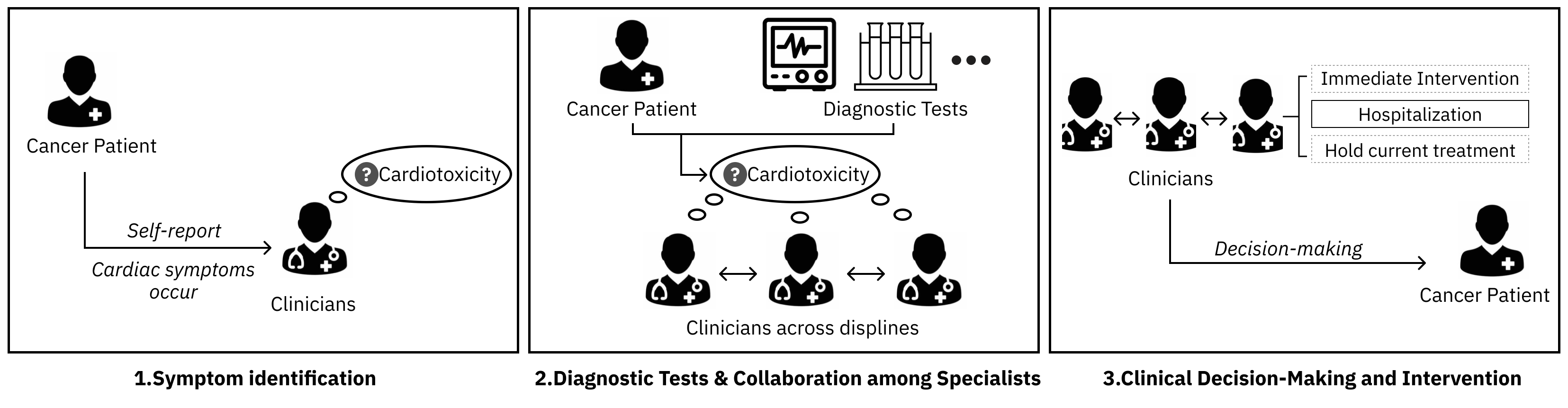}
    \caption{Current Workflow of Cardiotoxicity Diagnosis. First, cancer patients' self-reports or the presentation of acute cardiac symptoms lead clinicians to identify symptoms and suspect cardiotoxicity. Next, clinicians perform diagnostic tests and collaborate with other specialists to diagnose cardiotoxicity. Finally, clinical decisions regarding immediate intervention, treatment modifications, and continuous monitoring are made.}
    \label{fig:current-workflow}
\end{figure*}


After being diagnosed with cancer, patients receive the plan and schedule for their cancer treatment and follow-up visits. Cancer treatment-induced cardiotoxicity could happen any time after patients start their cancer treatment. 
From our interviews, we derive and organize the current cardiotoxicity risk management process into a pathway of three primary stages: 
symptom identification, diagnostic testing and collaboration with other specialties, and clinical decision-making and intervention, as shown in Figure~\ref{fig:current-workflow}.

\subsubsection{Symptom Identification}


The initial step in cardiotoxicity risk management is to identify any symptoms that may lead to cardiotoxicity.
The clinical complication of symptom identification can mainly be attributed to whether cardiotoxicity is symptomatic or asymptomatic and whether patients are hospitalized or not.

Cancer treatments often involve multiple cycles or sessions over a period, known as treatment cycles. Since treatment plans vary widely according to patient-specific factors, such as age, sex, medical history, cancer type, and cancer stage, the frequency of hospital visits differs. Usually, patients are seen at the beginning of each treatment cycle, with subsequent treatments within the cycle and follow-ups at the beginning of each new cycle. For example, P5 mentioned the varying frequencies of appointments for patients undergoing different treatments, \textit{``during chemo, it's every 3 weeks, and then radiation is every day.''}

When patients visit the hospital for different cancer-related purposes, such as cancer treatment or follow-up visits, they undergo routine monitoring, physical examinations, or treatment sessions. These clinical monitoring approaches are effective in identifying asymptomatic cardiotoxicity during hospitalization. 
For example, P5 recounted an instance in which a patient experienced bradycardia with premature ventricular contraction during a chemotherapy infusion, which was not immediately symptomatic but was detected through monitoring.
Patients can also self-report abnormal symptoms to clinicians for timely examinations and interventions during hospitalization.

However, as all participants mentioned, the current clinical practice heavily relies on patient self-reporting for symptom identification when patients are discharged from the hospital. 
There is a lack of an effective symptom identification strategy if the cardiotoxicity is asymptomatic and the patients are not hospitalized, which could lead to belated diagnosis and severe outcomes. 

The majority of self-reported symptoms related to cardiotoxicity include chest pain, shortness of breath, fatigue, palpitations, etc. 
For example, P2 noted that a patient came in complaining of chest pain shortly after starting a new treatment cycle. Patients use various methods to report their symptoms, often while at home. Many utilize the MyChart system, an online health management tool that provides personalized and secure access to portions of their medical records within the electronic medical record (EMR), to send messages to clinicians (P3, P4, P6).
For more severe symptoms, patients may call the office or the emergency line directly (P5, P6). 
For example, P4 described typical ways patients contact their health providers.

\begin{quote}
   \pquote{P4}{They usually contact us. they will either write a MyChart message, or if it's more severe than that, they will call 911, or if they just cannot wait, they'll go to the ER.}
\end{quote}

\subsubsection{Diagnostic Testing and Collaboration among Specialists}

Based on symptoms reported by patients or indications from physical examinations during hospitalization, clinicians usually conduct various diagnostic tests, such as EKGs, echocardiograms, etc., to check for determinative indicators of cardiotoxicity diagnosis. 
This process usually involves multiple departments and different clinical specialists to diagnose cardiotoxicity in a collaborative approach.
In this diagnostic process, the variety of tests and imaging are fundamental and critical. For example, P1 described a variety of different labs they conducted when immediate signs of cardiotoxicity were not evident,

\begin{quote}
    \pquote{P1}{We checked a variety of different labs... she didn't have troponin elevation, CK elevation... no visible EKG changes. The echocardiogram looked unchanged.}
\end{quote}



However, not all cardiotoxicity indications from examinations and self-reported symptoms are caused by cardiotoxicity--they may be attributed to a variety of causing factors, for instance, other diseases or drug-induced side effects. 
The clinicians need to gather more data via lab test results so that they can rule out false hypotheses about causing factors and determine whether the symptoms are related to cancer treatment-induced cardiotoxicity.

For example, P4 described a case in which a patient presents fatigue, and the clinical team performs a comprehensive work-up to identify the causing factors: 
\begin{quote}
    \pquote{P4}{We do the full work up. We do blood counts, we do other tests. We, for example, check for their thyroid. We check for other things. If everything else is negative, then we are sure that the disease is not progressing. The fatigue is not from the disease. Then we have every reason to believe the fatigue is from the drug.}
\end{quote}



Cancer treatment clinicians often need to consult with or refer patients to specialists, particularly cardiologists, for advanced diagnostic testing, management of confirmed cardiac events, or evaluation of atypical symptoms.
P1 elaborated on the process, explaining that while they, as cancer specialists, can technically order specialized tests like cardiac MRI, it is often done in consultation with specialists in related fields: 
\begin{quote}
    \pquote{P1}{Technically, we can still order cardiac MRI. I'll say mostly in our practice, we will consult it with either this rheumatologist or somebody in our cardinal oncology group will often be the one to support the ordering of this.}
\end{quote}

Similarly, P5 described the typical workflow to involve cardiologists for in-depth examinations when there is confirmed cardiac depression or a significant cardiac event: 
\begin{quote}
    \pquote{P5}{Typically, if we confirm that there is a cardiac like an actual cardiac depression or event, or anything, then we get cardiologists involved. If it turns out that there is no ejection fraction problem. There's no coronary issue or anything like that. Then typically, we're like, okay, then it's just the chemo(therapy), and they don't have to see the cardiologist then. That's at least what we typically do now.}
\end{quote}

Due to the substantial complication and high uncertainty of clinical situations, even in cases where patients have atypical symptoms that might not seem directly related to cardiac issues, they are often referred to cardiologists to ensure comprehensive evaluation and management, as P7 discussed: 
\begin{quote}
\pquote{P7}{It's more than even if they have atypical complaints like atypical chest pain not really related to the heart, they tend to get referred to cardiologists.}
\end{quote}

\subsubsection{Clinical Decision-Making and Intervention}
Once cardiotoxicity is suspected or diagnosed, clinical decisions are made regarding immediate interventions, medication adjustments or cessation, and continuous monitoring. 
For immediate intervention, hospitalization is often the first step considered to treat severe cases of cardiotoxicity. 
For example, P1 explained a typical hospitalization process of cardiotoxicity intervention:
\textit{``We admitted her to the hospital, where she then received higher doses of steroids and was considered for some of our clinical trials with cardiologists.''} 
During hospitalization, procedures such as cardioversion and rate control may be performed to stabilize the patient's condition before discharge.

Adjusting or discontinuing medications is another critical decision in managing cardiotoxicity in cancer patients. This process involves careful consideration of the patient's overall cancer treatment plan and the specific cardiac side effects that they are experiencing. It can involve stopping the offending agent, changing the treatment plan, or adding new medications to mitigate cardiac side effects. Stopping the offending agent is often the first line of action in this decision. As P4 mentioned, \textit{``I asked him to stop the medication right away,''} highlights the immediacy with which clinicians act to prevent further cardiac damage. 
Participants mentioned that simply stopping a medication is not always sufficient to control cardiotoxicity in some cases. 
To balance the need to treat primary cancer while preventing further cardiotoxic effects, clinicians may decide to continue the cancer treatment with new cardiac medications or alter the current treatment plan to a less cardiotoxic regimen. 
Such a decision-making process often involves the collaborative decision-making process by consulting cardiologists, as described by P7:


\begin{quote}
    \pquote{P7}{Sometimes we say, like, this is clearly related to the cancer treatment. We should not continue the cancer treatment and something else needs to be done. So that's when the cancer doctor then comes up with some other form of cancer treatment which might not be the best possible cancer treatment, but something close to best possible. But would not have the cardia(c) complication.}
\end{quote}

Nevertheless, compared with patients who only experience heart failure, cancer patients with cardiotoxicity are much more vulnerable, which leads to a much more complicated and higher stake clinical decision-making because their survival is intricately linked to their ongoing cancer treatment. 
For cancer patients, stopping or changing the best possible cancer treatment due to cardiotoxicity could dramatically affect their prognosis, as illustrated by P7:

\begin{quote}
    \pquote{P7}{...the question, then comes whether you need to continue cancer treatment or not, or whether you have to change the best possible cancer treatment that they get.}
\end{quote}

After managing the initial intervention phase of cardiotoxicity, decisions regarding continuous outpatient follow-up and monitoring become crucial to ensure ongoing assessment and effective management of the patient's cardiac health. 
P4 shared an example of sending a patient home with a heart monitor after an initial hospital workup, which was normal, to ensure continuous monitoring, \textit{``Everything was negative, everything was normal. But they sent him home with a monitor.''} For patients who have medication adjustments, regular follow-up appointments with clinicians are essential to understand the lasting impacts of cardiotoxicity and the effectiveness of interventions.

\subsection{Challenges in the Current Clinical Practices of Cardiotoxicity Risk Management}
\label{sec:findings-challenges}

The current clinical practices in diagnosing and monitoring cardiotoxicity in cancer patients face multiple challenges, which complicate the identification and management of cardiotoxicity, as symptoms can be ambiguous and testing methods inconsistent. Additionally, monitoring practices rely heavily on patient self-reporting, which is often inadequate due to various patient-related factors, further hindering timely diagnosis and treatment. 

\subsubsection{Challenges Related to the Diagnosis of Cardiotoxicity}




Diagnosis of cardiotoxicity in cancer patients presents significant challenges due to a variety of complications. We identified a few common clinical challenges, as stated by multiple participants, such as the ambiguity caused by overlapped toxicities, variability of symptoms, and the limitations of existing testing. 

\paragraph{Ambiguity Caused by Overlapped Toxicities}
Cancer treatments often come with a range of side effects that can impact various organs and systems within the body. The symptoms of cardiotoxicity could overlap those caused by cancer treatments. For example, chemotherapy treatments often have overlapping toxicities, making it challenging to distinguish between general chemotherapy side effects and true cardiotoxicity.
This overlap creates ambiguity and difficulty in diagnosis, as exemplified by P1, \textit{``if her side effect was rash. Rash is caused by those chemotherapies as well, so it is really hard to differentiate what is causing it.''} 

Symptoms such as fatigue and shortness of breath, which are not specific to any disease or treatment, are particularly difficult to interpret. 
These symptoms are common among cancer patients undergoing chemotherapy and may not necessarily indicate cardiac dysfunction, which further complicates identifying cardiotoxicity. P4 gave an example: 

\begin{quote}
    \pquote{P4}{There were more symptoms that he was having... A lot of those symptoms, fatigue, shortness of breath, you can have just from chemo(therapy)... With him specifically... the EKG was showing some changes as well... He ended up in the hospital feeling very short of breath... The ejection fraction was normal initially, but later it dropped to 30\%}     
\end{quote}

Besides, in some cases, the primary focus on symptoms related to the cancer itself can lead to the under-recognition of cardiotoxicity. Clinicians often prioritize cancer-specific symptoms, potentially overlooking cardiac issues. P2 illustrated this point with an example: \textit{``Cause this was a rectal cancer patient. We don't necessarily ask about chest pain normally because we're more focused on like the GI part of the disease.''}

\paragraph{Variability of Symptoms}

Participants noted that the manifestation of symptoms can vary significantly depending on the type of cancer being treated, even when the same chemotherapeutic agents are used. This variability complicates the diagnosis of cardiotoxicity, as certain symptoms may be more prevalent in one type of cancer compared to another. From our interviews, when participants recalled their recent encounters, we observed that various symptoms, such as edema, fatigue, chest pain, and arrhythmias, could be signs of cardiotoxicity. For instance, P1 described a patient on immunotherapy who experienced worsening edema and increased fatigue, whereas P2 noted a patient experienced chest pain likely due to coronary vasospasm. This diversity in symptom presentation underscores the complexity of diagnosing cardiotoxicity, as the same underlying issue can manifest differently depending on the patient's type of cancer and treatment regimen.

Adding to the complexity, some patients may not experience noticeable symptoms, making it difficult to identify cardiotoxic effects without close and continuous monitoring. For example, P4 recounted, \textit{``Because he was completely asymptomatic. He was sitting at home relaxing, and the monitor picked up this thing''}.

\paragraph{Limitations of Existing Testing}
Another critical aspect of these challenges is the limitation of existing testing. In the current practice of diagnosing cardiotoxicity related to cancer treatments, there is a lack of standardized testing protocols. During our interviews, clinicians pointed out that they often need to rely on a variety of diagnostic tests with specialist consultations for diagnosis. Even with the various tests, sometimes those tests may yield inconclusive results, adding complexity to the diagnosis process. P1  provided an illustrative example of this complexity, 

\begin{quote}
        \pquote{P1}{We checked a variety of different labs... She had liver enzyme increases... But she didn't have troponin elevation. CK elevation. Some of these things that we more often see with cardiotoxicity. She had no visible EKG changes... The echocardiogram looked unchanged}
\end{quote}

Furthermore, participants noted that, according to their clinical observations and encounters, some existing lab tests may not perform well in identifying subtle or early-stage cardiac changes. P1 remarked on this limitation, stating, \textit{``something that we weren't finding based on those labs, because those labs are, you know, somewhat limited sometimes in diagnosing some of these cardiac changes.''} This observation indicates the insufficiency of current diagnostic tools in capturing the full spectrum of cardiotoxicity. 

Interpreting lab results in the context of cardiotoxicity presents another challenge. For example, P5 gave an example of a patient having normal troponin levels or unchanged echocardiogram readings, which obscure the presence of cardiotoxicity. 

Additionally, the timeliness of the tests emerged as a crucial factor. Delays in obtaining necessary tests such as echocardiograms or EKGs can postpone diagnosis and subsequent treatment of cardiotoxicity, ultimately exacerbating patient well-being outcomes. 
P2 stated, \textit{``I had to wait. I had to wait to get the echo. I had to wait to get the stress echo. I had to wait to get the EKG, so basically, everything was kind of delayed.''}

\subsubsection{Challenges Related to the Monitoring of Cardiotoxicity}

Besides challenges regarding diagnosis, there remain significant challenges in effectively monitoring cardiotoxicity risks during cancer treatment or post-treatment. 

\paragraph{Timing and Methods of Monitoring}

The unpredictable onset of cardiotoxicity presents a significant challenge regarding when and how to monitor it. 

Cardiotoxicity can occur at any point during or after treatment, whether patients are hospitalized or at home. Patients can develop acute cardiotoxicity or chronic cardiotoxicity as P5 noted, \textit{``a lot of times they'll develop cardiotoxicity 6 months later, sometimes even a year or 2 later.''} 
Specific treatments may have more defined risk windows. For example, P5 indicated that patients who do not recover after the expected time frame for chemotherapy side effects, typically 7 to 10 days, might be experiencing more severe complications, including cardiotoxicity. 
Prolonged therapies, such as those involving inhibitors or immunotherapy, require extended monitoring periods. P1 described the unpredictable nature of side effects and the necessity of continuous monitoring throughout cancer treatment: 
\begin{quote}
    \pquote{P1}{
    these side effects can happen at any time during their treatment. 
    I think certain side effects are more likely to happen earlier in the course versus later. 
    This patient had already received almost 3 months of therapy...
    the problem is that these patients sometimes need to be on immunotherapy for a year... 
    So you kind of have to be on the lookout at all times for these issues.}
\end{quote}

Most monitoring of cardiotoxicity occurs within a clinical setting, especially during follow-up visits. However, the frequency and nature of these visits vary greatly depending on the type of cancer, treatment modalities, patient symptoms, and overall health. This variability further complicates the monitoring process, particularly for patients who are not hospitalized. For example, P1 mentioned that these patients may face challenges accessing timely follow-up care, especially if they live far from the centers. Additionally, the long intervals between follow-up visits can lead to delays in identifying and addressing cardiotoxicity. 

Also, monitoring protocols and guidelines vary by treatment type. As P7 mentioned, some cancer treatments have established protocols for regular echocardiograms, while others do not: 

\begin{quote}
    \pquote{P7}{
    So the cancer treatments which typically cause the heart function to get worse, there [are] guidelines or their cancer [has] guidelines, where they sometimes recommend getting an echo(cardiograms). but it's not there, and some cancer treatments which do not affect the heart function at all.}
\end{quote}

\paragraph{Existing Self-Reporting Tend to Downplay Symptoms}

While clinicians can closely monitor patients during hospitalization with timely examinations and interventions, monitoring outside of hospitalization heavily relies on patients reporting their symptoms, which faces several challenges.
The most cited challenge is patients either not reporting or downplaying the reporting of their symptoms despite encouragement to do so. 
For instance, as P1 mentioned, \textit{``There are a lot of patients that they are hesitant to call us in with symptoms.. I think some patients will downplay what's actually going on.''} 

Our participants provided valuable insights into the nuances behind why patients do not report or downplay symptoms. A significant factor is \textbf{the fear of bothering the clinicians} and appearing overly dramatic about their health concerns, as P1 stated, \textit{``I think it's also they don't want to feel like they're complaining too much.''} 
Another critical reason is patients' concern about \textbf{the impact on their treatment efficacy} due to a lack of health literacy. Patients may not understand the importance of reporting symptoms and often fear that reporting symptoms might lead to a reduction in their chemotherapy dose or a complete halt in treatment. P1 highlighted this concern, noting, \textit{``they're concerned about efficacy not being as good, or that I'm gonna stop their treatment because of toxicity''} 
Additionally, \textbf{the burden of managing additional medications} for symptom relief discourages patients from reporting their symptoms. Many patients find the regimen of taking multiple symptomatic medications, such as anti-nausea or anti-diarrhea drugs, to be cumbersome and disruptive to their daily lives. This inconvenience and the influence on their quality of life lead some patients to prefer to live with the symptoms rather than adding more medications to their routine.

\textbf{The expectation of delayed responses} from clinicians also contributes to patients' reluctance to report symptoms. Especially in resource-poor areas, patients often do not receive prompt attention to their reported symptoms, leading to a sense of futility in reaching out, especially if they have experienced slow responses in the past. 
\textbf{Geographical barriers and resource constraints} further exacerbate this difficulty. Some patients travel long distances to receive treatment, which limits their ability to seek timely care when symptoms arrive, and this situation is particularly pronounced in regions with limited local healthcare resources. 
\begin{quote}
    \pquote{P1}{A lot of our patients, especially here, they travel very long distances to come here for treatment. Oftentimes, they may call and be having an event but can't get in to see us the same day or even within the week... They don't have the resources, even if they did have changes in their symptoms, to get out here or be treated.}
\end{quote}

In addition to these reasons, there are limitations in the tools used for self-reporting. The existing survey tools for self-reporting may not be designed specifically to manage cardiotoxicity risk. As P5 mentioned, \textit{``We will use a a survey tool called Promise 10 to ask patients how they're feeling. But it's not specific to cardiotoxicity. ''}

The consequences of downplaying or not reporting symptoms in a timely manner can be severe. It can lead to significant delays in necessary diagnostic tests and medical interventions. As P2 pointed out, \textit{``you know she had the symptoms. Then she told me, like, essentially 4 or 5 days later. and then I had to wait. I had to wait to get the echo. I had to wait to get the stress echo. I had to wait to get the Kkg, so basically, everything was kind of delayed.''}

\subsubsection{Lack of Clinical Guideline}

Many of the previously discussed challenges are intrinsically linked to the absence of specific clinical guidelines for diagnosing and monitoring cardiotoxicity related to cancer treatment. This gap is particularly evident in the identification of symptoms and the conduction of appropriate diagnostic tests. 

P2 pointed out the general lack of focus on cardiac issues during routine oncology assessments:
\begin{quote}
    \pquote{P2}{In oncology clinics, we do a general review of systems, but there aren't a lot of specifics about cardiac issues}
\end{quote}

This oversight could partly be due to the rapidly evolving nature of cancer treatments and their side effects, which outpaces the development of comprehensive guidelines. For example, P1 emphasized the novelty and rapid development in this area, pointing out the lag in guidelines: 

\begin{quote}
    \pquote{P1}{So much of this that it's very new, and the guidelines for how we diagnose and identify it probably lag the actual clinical, you know, outcome and what we see in the clinic. }
\end{quote}

The absence of specific cardiotoxicity guidelines indicates certain levels of inconsistencies in care. P7 mentioned that some patients receive comprehensive cardiac monitoring, while others may not, depending on the treatment they are undergoing and the presence of specific guidelines, explaining, \textit{``so the cancer treatments which typically cause the heart function to get worse, there are guidelines or their cancer has guidelines, where they sometimes recommend getting echocardiograms. But it’s not there for some cancer treatments which do not affect heart function at all.''}

\subsection{Clinicians' Needs \& Perspectives Towards Potential Technology-Supported Solutions}
\label{sec:findings-needs}

In addressing the challenges of diagnosing or monitoring cardiotoxicity, participants expressed their needs and offered nuanced perspectives on the potential of digital health technologies, including advanced AI assistance.

\paragraph{Early Detection and Intervention}
One of the primary needs expressed by the participants is the ability for early detect and diagnose cardiotoxicity. The earlier symptoms are identified, the higher the chances of preventing serious complications.
P1 mentioned the early detection of serious side effects is crucial due to the overlapping toxicities of chemotherapeutic agents: 
\begin{quote}
    \pquote{P1}{sometimes they're very mild, like people can have a little bit of diarrhea. If it's small, then that's something that will pass. But if it's a true colitis that the patient is having that could be life threatening. So determining what those are earlier on helps.}
\end{quote}

While early identification of cardiotoxicity is critical, achieving this is often constrained by the availability of specialized resources and expertise. Particularly in less specialized clinical settings, early detection becomes more challenging, leading to potential delays in diagnosis and treatment. P1 pointed out:  

\begin{quote}
    \pquote{P1}{it's also very helpful for them (cancer specialists) to know when these patients need to be referred to somebody here or in a different center where they are evaluated by a cardio-oncology specialist... those patients have worse outcomes because we know delay in treatment with these cardiac events can be serious.}
\end{quote}

\subsubsection{Existing Supportive Tools}

Our participants mentioned several existing supportive tools they are aware of or have experienced using in clinical practice. 

One of these tools is the Chemotherapy Risk Assessment Scale for High-age Patients (CRASH), which predicts the risk of severe chemotherapy-induced hematologic and non-hematologic toxicity in older adults. This tool can potentially address the need for early detection and diagnosis. However, clinicians mentioned that the CRASH score requires manual calculation and is not automated, leading to a significant workload for clinicians. P6 mentioned that automating tools like CRASH would enhance their usability and efficiency. 

Another tool they have mentioned is the electronic Frailty Index (eFI), an automated tool that uses data from electronic health records (EHRs) to identify and categorize frailty in older adults. While it has been integrated into some systems, it is not designed specifically for cardiotoxicity risk management, which limits its direct applicability to address early detection and diagnosis. 

Additionally, AI technology, such as Deep 6 -- which aims to match patient cancer characteristics with clinical trials -- is mentioned. While this technology can reduce the time and effort needed for patient recruitment, it has not been widely implemented or commonly used among clinicians in real-world clinical settings. 

While the existing supportive tools have potential, they face different limitations or are not specifically designed for cardiotoxicity risk management. Given these inputs, we are interested in understanding what kind of technology clinicians envision to be helpful in addressing the need for early detection and diagnosis.

\subsubsection{Potentials for Remote Monitoring}

During our interviews, participants discussed the potential of telehealth technologies, such as wearable devices, for remote patient monitoring and early detection of symptoms. 

Current remote monitoring solutions in clinical settings often focus on patients who already exhibit certain cardiac risks. Temporary heart monitors, for instance, are frequently employed to detect abnormal heart rhythms in patients who report symptoms such as palpitations (P4, P5, P7).

Telehealth technologies could offer continuous monitoring capabilities, allowing clinicians to track patient health conditions between visits and pick up patients earlier, especially for patients who cannot frequently visit healthcare facilities. P1 explains that patients may experience symptoms episodically, making it difficult to capture those events during regular clinic visits:
\begin{quote}
    \pquote{P1}{I've had a couple of patients that are on potentially cardiotoxic agents, and they'll have episodes of almost passing out. But I'm not sure if, like that was correlated to something else. So, and obviously they're not often having them in the office when I'm seeing them, so that could be helpful.}
\end{quote}

Another significant potential of telehealth technologies, as mentioned by clinicians, is their ability to empower patients by involving them actively in their health management. As P1 mentioned, some cancer patients are often highly engaged in their care and motivated to adopt measures that can improve their overall health and well-being. 
Wearable telehealth technologies, such as smartwatches, could help patients gain better insight into their health status and make informed decisions about their care.

\subsubsection{Perspectives towards AI Technologies}

As clinicians raise the lack of automation as one significant limitation of existing risk management tools, we follow up on those comments to ask about their perspectives towards advanced AI technologies, such as chatbots and predictive risk scores. 

Clinicians are aware of ongoing advancements and potential future integrations of AI within EHR systems, which they are currently using in their day-to-day work; however, the current adoption of AI in clinical practice remains limited. For example, P4 expressed, \textit{``I don't know why I don't use any right now. I haven't felt the need to use any at this point.''} indicating a lack of a compelling need identified by some clinicians or a lack of experience using novel technologies. 

Our participants shared their perspectives on the role of AI. A prevailing theme is that AI aims to serve as an adjunct to human judgment, not a replacement. Many clinicians emphasize the importance of monitoring AI output to ensure that they augment rather than overshadow clinical expertise. P5 stated her position as \textit{``I know one thing that the physicians would definitely want to just make sure. That it's an adjunct versus the only decision maker.''} Similarly, P4 pointed out that \textit{``AI is supposed to be my assistant, not my replacement.''} 
Clinicians also emphasize that AI needs to provide actionable insight. P7 noted that while the influx of data could be useful, it can become overwhelming without the appropriate support to act on it. Then, the additional data becomes burdensome rather than beneficial.

\paragraph{Potentials of AI technologies}
Many clinicians (P1, P2, P4, P7) have mentioned the potential of AI to streamline the documentation process and summarize medical records. As P4 mentioned, \textit{``There's a lot of burnout in medicine, and the burnout is from all of this documentation that we do. And I see a huge role of AI in that documentation that will help with the burnout.''}, expressing the potential that AI could alleviate the administrative burdens contributing to burnout. 

Regarding AI predictive risk scores, clinicians see value in automating cardiotoxicity risk assessments. As the current risk tool is manual-intensive, P6 expressed that it would be nice if it could be done automatically.

\subsubsection{Concerns about Digital Health Solutions}

Clinicians have also expressed various concerns regarding using telehealth technologies and integrating AI technologies into their current workflow, including accuracy and reliability, alarm fatigue and data overload, and ethical and capacity issues. 

While the practical utility of AI-generated risk scores appears promising, nuanced perspectives have been discussed. A primary concern among clinicians is understanding what the risk scores mean. Clinicians emphasize the importance of clarity and validation in interpreting AI-generated risk scores, which revolves around the necessity of clear guidelines and reliable validation of the risk scores. P3 highlights the need for validation to ensure the accuracy and reliability of risk scores, \textit{``you need to validate and say, well, a risk score from one to 4, a very low risk for a heart attack.''} Another critical aspect discussed is how clinicians can effectively use these scores in their current workflow. Many of the participants (P1, P5, P6, P7) raise a critical question, \textit{``What do we do with it?''}, highlighting the need for actionable insights. They need concrete steps and recommendations on how to use the scores rather than just being presented with a number. 

When using general digital health technologies, alarm fatigue and data overload are significant concerns among clinicians. Alarm fatigue could be stemmed from an excessive number of alerts generated by remote monitoring technologies, AI systems, and electronic health records. Clinicians express frustration with the constant interruptions caused by non-essential alerts. This could lead to healthcare providers becoming desensitized to alarms, resulting in potential neglect of critical alerts. P4 expressed this sentiment: 
\begin{quote}
    \pquote{P4}{But you know, doctors also get annoyed about being sent every little detail. We don't want to be woken up by every little thing, so it should not be annoying like that.}
\end{quote}
Data overload could be another critical issue. The sheer volume of data can become overwhelming without appropriate mechanisms to distill it into actionable insights. Organizing and finding relevant information from all the data could add extra burden to clinicians. 

Clinicians have expressed their ethical concerns regarding the use of technologies. One of the primary ethical concerns revolves around the responsibility of monitoring the ``new'' data. Clinicians express apprehension about who will be responsible for continuously overseeing the data produced by remote monitoring devices or AI systems and ensuring timely interventions based on the data. P1 articulated the ethical dilemma:
\begin{quote}
     \pquote{P1}{There's always the issue of ethics... if something like that happens like, when do we find out about this? Is it ethical like for us to wait to see this data every time they come in? Versus? How are we gonna have somebody monitoring this data all the time, like the next man call, who's gonna be monitoring this. }
\end{quote}
This concern is also related to the capacity to manage the influx of data and alerts, which needs the necessary resources and infrastructure to respond effectively. Liability is another significant concern among clinicians, particularly regarding the actions taken based on AI-generated recommendations. Under the context of the AI intervention in remote monitoring, P5 raised her concern: \textit{``Do we have the chatbot recommend an intervention such as going to the ER, or is it more so that the chatbot then gets flagged that contact the nurse or doctor now?''}

%% file: sections/5-discussion.tex
\section{Discussion}

The diagnosis and monitoring of cardiotoxicity in cancer patients is a critical area of concern, given the complexity and variability of symptoms, as well as the life-threatening implications, which also becomes a typical example of high-stake, high uncertainty, and continuously evolving clinical scenarios.
In Section~\ref{sec:findings}, we analyzed the challenges and intricacies in the current clinical practices, explored the role of technologies, and discussed the concerns raised by clinicians. 
In the following, we discuss our findings in relation to the existing literature and propose design considerations for future research and technological advances to support the workflow of clinicians. 
More importantly, we delve deeper into problems rooted in the underlying phenomena of constantly evolving clinical settings and the prevalent critical collaborative paradigms in clinical experts' workflows.

\subsection{Design Considerations}






A major observation regarding clinicians' needs in current cardiotoxicity management, as illustrated in Section \ref{sec:findings-needs}, is the emphasis on early detection of cardiotoxicity. Clinicians noted that delays in diagnosis can lead to severe complications. Reflecting on the participants' perspectives on the potential of technologies that could be used to enhance early detection and current cardiotoxicity risk management, we propose the following potentials and design considerations regarding the functionality, alignment with clinical research, and accessibility of these technologies. 


\textit{Potential for Remote Monitoring Technologies: Continuous Monitoring.} Remote monitoring technologies present a valuable solution for continuous patient care after discharge. Patients benefit from systematic monitoring and timely intervention during hospitalization. However, the risk of complications may increase after discharge. Previous studies have shown that telehealth platforms and wearable devices can facilitate continuous monitoring, allowing clinicians to track patient health conditions in real time~\cite{hardcastle2020fitbit, gresham2018wearable, cox2018use}. Our participants have discussed the potential of remote monitoring technologies in continuous monitoring and empowering patients to self-manage their well-being. Prior studies have utilized wearable devices to monitor heart rate and oxygen saturation in cancer patients, demonstrating their effectiveness in providing continuous, real-time data that aids patient management~\cite{lachenmeier2022home, gresham2018wearable, cox2018use}. 

\textit{Potential for AI Technologies: Risk Prediction.}
AI technologies offer powerful capabilities for risk prediction, which is essential for the early detection of cardiotoxicity. For instance, recent research~\cite{yagi2024artificial} has demonstrated how AI can be trained with clinical, chemotherapy, and echocardiographic parameters to predict cancer therapy-related cardiac dysfunction with high accuracy. This predictive capability allows for proactive management and early intervention, potentially preventing severe cardiotoxicity

\textit{Potential for AI Technologies: Automating Documentation Tasks.}
Automating routine documentation tasks through AI can significantly enhance clinical efficiency, supporting early detection efforts. Our participants noted that many clinical tasks require significant manual effort, and automation of some repetitive and low-risk work could significantly enhance work efficiency. For example, AI systems can handle data entry and initial risk assessments, ensuring accurate and timely clinical data recording. This can free up clinicians’ time, allowing them to focus on patient interactions and complex decision-making processes, improve efficiency, and reduce the likelihood of human error in routine processes.

Despite these potentials, we propose the following design considerations that need to be considered in designing and implementing these AI technologies to ensure they effectively meet clinical needs.

\textit{Functionality and Responsibility Allocation.} Our participants mentioned that the design of remote monitoring technologies should carefully consider their functionality and allocation of responsibilities. For example, if a chatbot is designed to check in with patients daily about their symptoms, it needs to handle the nuances of symptom reporting and follow-up actions. A critical question could be whether the chatbot should provide suggestions when patients report certain symptomatic symptoms, and if so, what kind of suggestions it should offer and how it should handle emergency situations.  These suggestions need to be perceived as advice rather than definitive clinical decisions to avoid miscommunication and potential liability issues. Additionally, determining who is responsible for monitoring the continuous flow of data generated by these devices is essential. Mismanagement or miscommunication of health information could lead to severe consequences. 

\textit{Sociotechnical Considerations.} Implementing remote monitoring technologies in outpatient settings presents several challenges related to sociotechnical factors. Patients' varying technological and health literacy levels, socioeconomic disparities, accessibility issues, and lifestyle factors can all impact the effective use of these technologies. When designing remote monitoring technologies, these factors must be considered to ensure they are user-friendly and accessible to patients of all backgrounds. 

\textit{Role and Purpose of Technologies.}
Our participants have emphasized the importance of clarifying the role and purpose of these technologies. Technologies should function as adjuncts to human decision-making, augmenting rather than replacing clinical judgment and providing actionable support and insights. It is crucial that these technologies do not overburden healthcare providers but instead integrate seamlessly into existing clinical workflows. 

\textit{Alignment with Clinical Research.}
For remote monitoring and AI technologies to be effective, their development must align with clinical research. Identifying specific parameters that need monitoring and determining indicators of cardiotoxicity ensures that these technologies are grounded in clinical evidence. Collaborative efforts among clinicians, patients, and HCI researchers are essential to establish critical values and alert thresholds that ensure the safety and efficacy of these technologies.

\subsection{The Significance of Collaborative Paradigm in Clinical Experts' Workflows}

Throughout the interviews, we have observed the significance of collaborative paradigms in managing cancer-related cardiotoxicity. Oncologists frequently collaborate with other specialists, particularly cardiologists, to develop diagnosis and treatment plans for cardiotoxicity. 
This is a common clinical scenario in which clinicians confronted with complex cases involving comorbidities rely on specialists in different fields of expertise to provide professional and nuanced advice. This collaborative effort is crucial for providing comprehensive care but poses significant challenges.

The key among these challenges could be the timing and nature of referrals and the modalities of collaboration. Deciding the optimal time for a referral is critical; a delay or premature referral can significantly impact patient outcomes. The nature of the referral itself -- whether a one-time consultation or an ongoing collaborative effort -- also plays a significant role. The modalities of collaboration involve the methods and channels through which specialists communicate and share information. 

For example, regarding the challenge of collaboration modalities, consider a scenario in which a patient undergoing cancer treatment begins to show early signs of cardiotoxicity, as identified by the cardiologist. In this situation, the cardiologist may recommend stopping the cancer treatment regimen to reduce the risk of further cardiac damage. However, this adjustment might not be optimal from an oncological perspective, potentially reducing the effectiveness of cancer treatment. Although all specialists are committed to finding the best possible outcome for the patient, their differing priorities and perspectives can lead to conflicting recommendations. Balancing these conflicting opinions could be a significant clinical challenge in this collaborative paradigm. This may involve sharing information and expertise, negotiating, and sometimes compromising to achieve a consensus that serves the patient’s best interests.

In addition, patients play a critical role in this collaborative process. To ensure truly patient-centered care, patients' needs, concerns, and preferences should be taken into account in the decision-making process. Their active participation is essential to avoid the power dynamics in medical consultations that might marginalize patient voices. Effective communication with patients about the potential risks and benefits of different treatment options is not just about conveying information but involves empathetic engagement that respects their autonomy and lived experiences. 

The timeliness and effectiveness of communication among specialists and patients are critical to collaborative decision-making. We advocate that future HCI research should identify when the clinician team needs collaborative decision-making, what each specialist values, how they communicate and collaborate in such clinical settings, and the challenges that arise during these collaborations. Additionally, the design, development, and evaluation of any technology-mediated tools to support clinical decision-making need to be grounded in the actual collaborative workflow. To achieve this, these tools should be evaluated in real-world collaborative decision-making, ensuring they support the nuanced and multi-faceted nature and incorporate patient engagement and agreement. 

\subsection{The Evolving Nature of Clinical Settings}
Our findings underscore the rapidly evolving nature of cancer treatments and clinical practices in diagnosing and monitoring cardiotoxicity. This fluidity is critical for research in this field, as new therapies are continuously developed, validated, and clinically adopted, necessitating constant adaptation in clinical workflows.

The introduction of novel cancer treatments often requires clinicians to learn and adapt to new workflows and toxicity profiles quickly. The workflow of diagnosing and monitoring cardiotoxicity is particularly fluid because of its close association with different clinical specialties, which requires clinicians to take careful considerations from multiple perspectives to make one or more clinical decisions. For example, chemotherapy regimens have evolved significantly over the years. Historically, the focus might have been on the management of gastrointestinal comorbidities, but with advances in treatments that prolong patient lifespans, new complications, such as cardiotoxicity, have emerged. 

Looking forward, the development of novel therapies with less known toxicities may reveal previously overlooked symptoms and new toxicities that do not exist in existing therapies, necessitating continuous clinical observations and studies. For example, traditional chemotherapies pose severe toxicity to the liver or kidneys~\cite{malyszko2020link, lefebvre2017kidney, munoz2017radiation, shanholtz2001acute}, whereas novel treatments, with much less damage to these organs, were found to cause acute and long-term cardiotoxicity -- a change in the primary cause of death from time to time as the treatment evolves.
Future therapies may also shift focus to other emerging side effects, requiring clinicians to remain vigilant and adaptable. This continuous evolution in treatment modalities means that clinicians must always be prepared to adapt their practices and workflows to accommodate new therapeutic advances and the resulting changes in patient conditions.

Furthermore, this evolving nature of treatments and clinical workflows often leads to a lag in the development of clinical guidelines, monitoring tools, and strategies. This lag is particularly significant in managing cardiotoxicity in cancer patients, highlighting a substantial gap in the availability of specific guidelines~\cite{jurcut2008detection,shelburne2014cancer}. Our participant's observations reinforce this, noting the slow pace of developing and implementing comprehensive guidelines that match the rapid advances in cancer treatment.
Clinicians play a crucial role not only as decision-makers but also as information collectors and observers. Their clinical experiences and observations provide invaluable insights that can inform researchers, HCI designers, and policymakers about emerging challenges and trends in the field. This feedback loop is essential for identifying new issues and developing solutions to address them effectively.




The changing nature of treatment modalities and side effects necessitates a corresponding evolution in clinical decision-making and workflows, such as adjustments in the monitoring period, diagnostic tests, and indicative biomarkers. When designing supportive tools under this evolving nature, HCI researchers need to be vigilant and proactive in identifying any changes in the clinical decision-making process and ensure that our understanding is well aligned with these shifts.

\subsection{Limitation \& Future Work}
Our work has several limitations. 
Firstly, our study population was relatively limited, consisting of only seven physicians, which is similar to the number of expert participants in previous studies~\cite{sepsis, cai2019hello, verma2023rethinking}. 
Future research should include a more diverse group from multiple hospitals. 
Despite this limitation, using this professional context for our research and adhering to established interview research principles~\cite{fusch2015we}, we have reached a saturation point, which allowed us to gain a comprehensive understanding of physicians' practices and challenges related to cardiotoxicity, as well as their emerging needs and expectations for future technologies.

Secondly, the variability of cardiotoxicity among cancer types, treatments, and individual patient characteristics presents a significant challenge. Future work aiming to develop a deployable system will need to start with a small, homogeneous group of patients -- such as those with the same cancer type undergoing similar therapies within similar medical backgrounds. This focused approach will help in creating more precise and tailored solutions that can later be expanded to diverse patient populations.

Moreover, while our focus on cardiologists and oncologists is crucial for understanding the intersection of cancer treatment and cardiotoxicity, we recognize the importance of involving professionals from different disciplines to gain a holistic view, such as emergency room doctors and nurse practitioners. Additionally, it is essential to engage patients in the research process to understand their experiences and expectations, and incorporate their needs in the design of technologies.







%% file: sections/6-conclusion.tex
\section{Conclusion}

This study sheds light on the critical challenges and opportunities in managing cancer treatment-induced cardiotoxicity. By engaging with clinical experts through semi-structured interviews, we have delineated the multifaceted decision-making process involved in cardiotoxicity risk management, emphasizing the complexities of symptom identification, diagnostic testing, and intervention strategies. Our findings reveal significant challenges, including the variability of symptoms, overlapping toxicities, and the absence of standardized protocols, which complicate timely and accurate diagnosis. 
Additionally, the difficulties in monitoring, due to patients' tendencies to underreport symptoms and the challenges of long-term follow-up, highlight the need for innovative solutions.
Our investigation of the potential of digital health technologies demonstrated that clinicians see value in early detection tools, remote monitoring, and other smart technologies. The responsible design and development of these technologies must be seamlessly integrated into existing clinical workflows and address clinicians' concerns about usability and reliability.

We provide several design considerations for future technology development, emphasizing the importance of clinician-centered approaches. These include ensuring early detection capabilities, facilitating accurate symptom monitoring, and supporting collaborative decision-making processes. Our findings contribute to the broader understanding of how digital health tools can be designed and implemented to enhance cancer treatment-induced cardiotoxicity risk management, ultimately improving cancer patient care outcomes.